\documentstyle[aps,prl,multicol,epsfig]{revtex}

\begin{document}

 \newcommand{\beq}{\begin{eqnarray}}
 \newcommand{\eeq}{\end{eqnarray}}
 \renewcommand{\thefootnote}{\fnsymbol{footnote}}

 \title{Nonextensive statistical mechanics and economics}

 \author{
 Constantino Tsallis and Celia Anteneodo \\
 }
 \address{
 Centro Brasileiro de Pesquisas F\'\i sicas \\
 Rua  Xavier Sigaud 150, 
 22290-180 Rio de Janeiro, RJ, Brazil 
 }

 \author{
 Lisa Borland and Roberto Osorio \thanks{tsallis@cbpf.br, celia@cbpf.br, lisa@sphinx.com, roberto@sphinx.com}\\
 }
 \address{
 Evnine-Vaughan Associates, Inc. \\
 456 Montgomery Street, Suite 800, San Francisco, CA 94104-1279, USA 
 } 
 \date{\today}

 \maketitle
 \begin{abstract}
 Ergodicity, this is to say, dynamics whose time averages coincide with ensemble averages, 
 naturally leads to Boltzmann-Gibbs (BG) statistical mechanics, hence to standard thermodynamics. 
 This formalism has been at the basis of an enormous success in describing, among others, 
 the particular stationary state corresponding to thermal equilibrium. There are, however, vast 
 classes of complex systems which accomodate quite badly, or even not at all, within the BG formalism. 
 Such dynamical systems exhibit, in one way or another, nonergodic aspects. In order to be able to 
 theoretically study at least some of these systems, a formalism was proposed 14 years ago, 
 which is sometimes referred to as nonextensive statistical mechanics.  
 We briefly introduce this formalism, its foundations and applications. Furthermore, we provide 
 some bridging to important economical phenomena, such as 
 option pricing,  return and volume distributions observed in the financial markets, 
 and the fascinating and ubiquitous concept of risk aversion.  
 One may summarize the whole 
 approach by saying that BG statistical mechanics is based on the entropy $S_{BG}=-k \sum_i p_i \ln p_i$, 
 and typically provides {\it exponential laws} for describing stationary  states and basic time-dependent  
 phenomena, while nonextensive statistical mechanics is instead based on the entropic 
 form $S_q=k(1-\sum_ip_i^q)/(q-1)$ (with $S_1=S_{BG}$), and typically provides, for the same type 
 of description, (asymptotic) {\it power laws}.

 \end{abstract}

\vspace*{1cm}



 Connections between dynamics and thermodynamics are far from being 
 completely elucidated. Frequently, statistical mechanics is presented as a 
 self-contained body, which could dispense dynamics from its formulation. 
 This is an unfounded assumption 
 (see, for instance, \cite{egdcohen} and references therein). Questions still remain open even 
 for one of its most well established equilibrium concepts, namely the Boltzmann-Gibbs (BG) factor 
 $e^{-E_i/kT}$, where $E_i$ is the energy associated with the $ith$ microscopic state of a 
 conservative Hamiltonian system, $k$ is Boltzmann constant, and $T$ the absolute temperature. 
 For example, no theorem exists stating the necessary and sufficient conditions for the use of 
 this celebrated and ubiquitous factor to be justified. In the mathematician F. Takens' words \cite{takens}: 
  
 {\it The values of $p_i$ are determined by the following 
 dogma: if the energy of the system in the $i^{th}$ state is $E_i$ and 
 if the temperature of the system is $T$ then: 
 $p_i=\exp\{-E_i/kT\}/Z(T)$, where $Z(T)=\sum_i \exp\{-E_i/kT\}$, (this 
 last constant is taken so that $\sum_ip_i=1$). This choice of $p_i$ is called {\it Gibbs distribution}. 
 We shall give no justification for this dogma; even a physicist like Ruelle 
 disposes of this question as ``deep and incompletely clarified".}

 One possible reason for this essential point having been poorly emphasized is that when dealing with 
 short-range interacting systems, BG thermodynamical equilibrium may be formulated without much referring 
 to the underlying dynamics of its constituents. One rarely finds in textbooks much more than a quick mention 
 to {\it ergodicity}. A full analysis of the microscopic dynamical requirements for ergodicity to be ensured 
 is still lacking, in spite of the pioneering studies of N. Krylov \cite{krylov}. In his words:

 {\it In the present investigation, the notion of ergodicity 
 is ignored. I reject the ergodical hypothesis completely: it is both 
 insufficient and unnecessary for statistics. I use, as starting point, 
 the notion of motions of the mixing type, and show that the essential 
 mechanical condition for the applicability of statistics consists in 
 the requirement that in the phase space of the system all the regions 
 with a sufficiently large size should vary in the course of time in 
 such a way that while their volume remains constant -- according to 
 Liouville's theorem -- their parts should be distributed over the whole 
 phase space (more exactly over the layer, corresponding to given values 
 of the single-valued integrals of the motion) with a steadily 
 increasing degree of uniformity. [...] The main condition of mixing, 
 which ensures the fulfillment of this condition, is a sufficiently rapid 
 divergence of the geodetic lines of this Riemann space (that is, of the 
 paths of the system in the $n$-dimensional configuration space), 
 namely, an exponential divergence (cf. Nopf $^1$).}

 Another possibly concomitant reason no doubt is the enormous success, since more than one century, 
 of BG statistical mechanics for very many systems. However, as complex systems become more and more 
 prominent in the front line of research [by {\it complex} we mean here the presence of at least one, 
 typically more, of the following features: long-range interparticle interactions, long-term microscopic 
 or mesoscopic memory, (multi)fractal nature of a pertinent subset of phase-space where the system remains 
 long time or forever, small-world or scale-free networking; see \cite{gell-mann} for a comprehensive 
 introduction to the subject], this fundamental issue starts being revised (see \cite{baranger} and 
 references therein).

 Indeed, a significant amount of systems, e.g., turbulent fluids (\cite{turbulentbeck,turbulentarimitsu} 
 and references therein), electron-positron annihilation \cite{bediaga}, cosmic rays \cite{cosmic}, 
 economics \cite{economics1,economics2,economics3}, motion of {\it Hydra viridissima} \cite{hydra}, 
 kinetic theory \cite{ademir}, classical chaos \cite{chaos1}, quantum chaos \cite{chaos2}, 
 quantum entanglement \cite{entanglement}, long-range-interacting many-body classical Hamiltonian 
 systems (~\cite{longrange} and references therein), internet dynamics \cite{internet}, and others, 
 are known nowadays which hardly, or not at all, accomodate within BG statistical mechanical concepts. 
 Systems like these have been successfully handled with the functions and concepts which naturally emerge 
 within nonextensive statistical mechanics \cite{tsallis1,tsallis2,tsallis3}.

 The basic building block of nonextensive statistical mechanics is the nonextensive entropy \cite{tsallis1}
 \beq
 S_q \equiv k\frac{1-\sum_{i=1}^{W} p_i^q}{q-1} \qquad (q \in {\cal R}).
 \label{sq}
 \eeq
 The entropic index $q$ characterizes the statistics we are dealing with; $q=1$ recovers the usual 
 BG expression, $S_1=-k\sum_{i=1}^W p_i \ln p_i$. We may think of $q$ as a biasing parameter: 
 $q<1$ privileges rare events, while $q>1$ privileges common events. Indeed, $0<p<1$ raised to a power 
 $q<1$ yields a value {\it larger} than $p$, and the relative increase $p^q/p=p^{q-1}$ is a {\it decreasing} 
 function of $p$, i.e., values of $p$ closer to 0 (rare events) are benefited. Correspondingly, for $q>1$, 
 values of $p$ closer to 1 (common events) are privileged. Therefore, the BG theory (i.e., $q=1$) is the 
 unbiased statistics. A concrete consequence of this is that the BG formalism yields {\it exponential} 
 equilibrium distributions (and time behavior of typical relaxation functions), whereas nonextensive 
 statistics yields (asymptotic) {\it power-law} distributions (and relaxation functions). Since the BG 
 exponential is recovered as a limiting case, we are talking of a {\it generalization}, not an alternative.

 The optimization of the entropic form (1) under appropriate constraints \cite{tsallis1,tsallis2} yields, 
 for the stationary state, the following distribution of probabilities:

 \beq
 p_i=\frac{[1-(1-q)\beta_q (E_i-U_q)]^{1/(1-q)}}{Z_q} \;,
 \eeq 
 where
 \beq
 Z_q \equiv \sum_j [1-(1-q)\beta_q (E_j-U_q)]^{1/(1-q)} \;,
 \eeq
 with
 \beq
 U_q \equiv \frac{\sum_j p_j^q E_j}{\sum_jp_j^q} \;,
 \eeq
 and
 \beq
 \beta_q \equiv \frac{\beta}{\sum_j p_j^q} \;,
 \eeq
 $\beta$ being the optimization Lagrange parameter associated with the generalized internal energy $U_q$. 
 Equation (2) can be rewritten as
 \beq
 p_i \propto [1-(1-q)\beta^\prime E_i]^{1/(1-q)} \equiv e_q^{-\beta^\prime E_i} \;,
 \eeq
 where $\beta^\prime$ is a renormalized inverse ``temperature", and the {\it $q$-exponential function} 
 is defined as $e_q^x \equiv [1+(1-q) x]^{1/(1-q)}=1/[1-(q-1) x]^{1/(q-1)}$ (with $e_1^x=e^x$).  
 This function replaces, in a vast number of relations and phenomena, the usual BG factor. 
 In particular, the ubiquitous Gaussian distribution $\propto e^{-ax^2}$ becomes generalized 
 into the distribution $\propto e_q^{-a_q x^2} = 1/[1+(q-1) a_q x^2]^{1/(q-1)}$ 
 (fat-tailed if $q>1$, and compact support if $q<1$).

 The use of the concepts and methods of statistical mechanics and thermodynamics in economics has long proved to 
 be a fruitful one (\cite{stanley} and references therein). It has even created a specific field of 
 research, whose name -- {\it Econophysics} -- has been proposed by Stanley a few years ago. Let us now 
 briefly review three recent applications of the ideas associated with nonextensive statistical 
 mechanics to phenomena in economics, namely a simple trading model which takes into account risk 
 aversion \cite{economics1}, a generalization of the Black-Scholes equation for pricing options 
 \cite{economics2}, and a phenomenological description of distributions of returns and volumes 
 in the real market \cite{economics3}. 

\vspace*{3mm}
 \noindent
 {\it Application to risk aversion:} 
\vspace*{3mm}

 The works by Kahneman, Tversky and others \cite{aversion} have put into quantitative 
 evidence the generic and enormous importance of risk aversion 
 (when one expects to win) and risk seeking (when one expects to lose) in trading. 
 The biased averages which naturally emerge in nonextensive statistical mechanics 
 ressemble those proposed in {\it prospect theory}, and constitute a tool that can be used 
 to make \cite{economics1} simple, and relatively realistic, models for a stock exchange or 
 similar forms of trading. 
 In fact, in reference \cite{economics1},  
 an automaton is introduced which simulates monetary transactions among operators 
 with different attitudes under risky choices. 
 Each operator is characterized by a parameter $q$ which measures 
 his(her) attitude under risk; this index $q$ is the 
 entropic one which plays a central role in nonextensive statistical 
 mechanics. Elementary operations are of the standard type used in 
 hypothetical choice problems that exhibit risk aversion \cite{aversion}. 
 By following the asset position of the operators, it is possible 
 to conclude on the consequences of each particular attitude on the 
 {\em dynamics} of economic operations.

\vspace*{3mm}
 \noindent
 {\it Application to option pricing:} 
\vspace*{3mm}

 The celebrated Black-Scholes equation provides, in an explicit manner, the
 prices of options. Its basic formulation assumes the presence of a Gaussian 
 noise, which simplifies the mathematical treatment on one hand, but yields 
 results which diverge from those observed in real markets.
 One of us \cite{economics2} has recently generalized this equation by assuming 
 non-Gaussian fluctuations, evolving anomalously in time according to a
 nonlinear Fokker-Planck equation. Such a noise process can be modeled
 as the result of a standard Gaussian process with  statistical feedback, of 
 the type discussed in
 \cite{pdependent}. The degree of feedback is characterized by the 
 nonextensive entropic index  $q$; if $q > 1$, then rare events will lead to 
 large fluctuations, whereas more common events will result in more moderate 
 fluctuations.
 This model of stock returns is consistent with empirical observations 
 of the distribution of returns. It is used as  a model of stock price 
 fluctuations, from which the fair value of options on the underlying stock 
 can then be derived\cite{economics2}. 
 The results are appreciably more realistic than the 
 usual Black-Scholes equation which is recovered as the $q=1$ particular case.

 This extended model, which preserves the practical advantage of having 
 explicit closed-form solutions, has proved to be a quite satisfactory one. 
 The theoretical option prices predicted by this model, using $q = 1.4$ which  
models well the underlying returns distribution, are in very good agreement 
 with empirically observed  option prices.  
 Let us look at the example of European call options. Such options are the  
 right to sell, at expiration time $T$, the underlying stock at an agreed  upon 
 price $K$, called the strike price.  
 Whereas the standard Black-Scholes equation must use a different value of 
 the volatilty $\sigma$ for each value of the option strike price in order 
 to produce theoretical values which match empirical ones, the $q=1.4$ model 
 uses {\it just one value of $\sigma$ across all strikes}. 
 For the standard Black-Scholes model, a plot of $\sigma$ versus the strike 
 $K$ forms a convex curve known as the volatility smile. We can use the $q=1.4$ 
 model with a fixed $\sigma$ to produce theoretical option prices, and we can 
 then find those $\sigma$ which the standard Black-Scholes model must use in 
 order to produce option prices which match the $q=1.4$ ones. A comparison of 
 such a theoretically obtained volatilty smile with the one observed from 
 market data will  reflect how closely the $q=1.4$ model fits real option  
 prices. As can be seen in Figure~1, where the option smile for Japanese Yen 
 futures is studied, there is a very close agreement.

\vspace*{3mm}
 \noindent
 {\it Application to financial returns and volumes:} 
\vspace*{3mm}

In the last few years, the easier availability of databases with detailed transaction
histories has enabled an increasing interest in the empirical study of distributions of
high-frequency financial variables (see, e.g. \cite{liu99,ple99,gop00,gop00volu,abde01}).
We are concerned in this section with two types of single-stock variables defined over fixed
time intervals: {\it returns}, defined here as logarithmic relative price changes, and
{\it volumes}, i.e., numbers of shares traded.

When {\it normalized returns} (after, for each stock, subtracting their mean and dividing
by their standard deviation) are measured over intervals of a few minutes, their distributions
are very well fitted by {\it q}-Gaussians with $q \simeq 1.4$. For long
periods (months or years), it seems that values of $q$ that approach the
Gaussian limit ($q = 1$) provide an adequate description. Results for
high-frequency normalized returns are presented in Figs. 2 and 3 \cite{economics3} for the  10 top-volume
stocks in 2001 in each of the two largest U.S. exchanges: the New York Stock Exchange (NYSE)
and the National Association of Security Dealers Automatic Quotation (NASDAQ).

Similar results are obtained for the corresponding distributions of {\it normalized volumes}
(divided by their means), by fitting them with {\it q}-exponentials multiplied by a simple power
of the normalized volume. Typical examples are presented in Figs. 4 and 5 \cite{economics3}. 
Previous studies
have shown that  volumes present a power-law behavior at high values \cite{gop00volu},
while {\it volatilities} present both a power-law behavior at high values \cite{liu99} and a
log-normal behavior at smaller values \cite{liu99,abde01}. Our curves are reminiscent of the
behavior of volatility curves, but we suggest, in addition, a new power-law regime at low volumes
and propose a functional form that unifies the description of the three regimes.
A model whose solution would be this type of distribution would of course be very welcome.

\vspace*{3mm}

With regard to the specific distributions associated with returns (Figs. 2
and 3), and similar ones ubiquitously observed in nature, a relevant comment
is appropriate. The two basic versions of the Central Limit Theorem (CLT)
are the following: (1) A convoluted distribution with a finite second moment
approaches, in the limit of $N \rightarrow \infty$ convolutions, a Gaussian
attractor. (2) A convoluted distribution with a divergent second moment
approaches, in the same limit, a L{\'e}vy distribution $L_{\gamma}(x)$ (with $2 >
\gamma > 0$). If we consider, for example, the {\it q}-Gaussian distributions indicated
above, then (a) if $q < 5/3$, the attractor is a Gaussian (i.e., $\gamma = 2$),
(b) if $5/3 < q < 3$, the attractor is a L{\'e}vy distribution, satisfying 
$q=(3+\gamma)/(1+\gamma)$, and (c) if $q = 5/3$ we have a marginal case involving
logarithmic corrections. The index $q$ cannot exceed 3 if the norm is to remain finite.

Can we then conclude that {\it all} stationary distributions found in nature
should be of either the Gaussian or the L{\'e}vy type? By no means! It would be
so only if the unique dynamics emerging in nature were the convolution
dynamics, i.e., with no memory across successive steps. In other words, it
would be so if the Laplacian term of the associated Fokker-Planck equation
in free space were either a standard second derivative or a fractional
derivative. In the first case, the solution is a Gaussian; in the second
case (for a derivative of order $\gamma$), it is a L{\'e}vy distribution.
(We assume throughout this discussion the standard first time derivative.)

Actually, much more complex and rich dynamics clearly exist in nature, for
example, those associated with a variety of nonlinear Fokker-Planck
equations involving nontrivial correlations, multiplicative noise and other
effects \cite{nonlinearFP,bologna}, whose exact solutions are {\it q}-Gaussians. 

This remark might seem trivial (it has in fact been already 
 mentioned in \cite{bologna}), but this point has
been overseen by some authors (e.g., \cite{zanettemontemurro}). To clarify this, recall that the
simple convolution CLT (in both the traditional and the L{\'e}vy-Gnedenko versions) only
allows, in the large N limit, for $q = 1$ (Gaussians) and $q > 5/3$ (L{\'e}vy
distributions). It does not allow in any way for fat-tailed distributions
associated with $1 < q < 5/3$. But it happens that, in many cases, complex systems in
both nature and the social sciences seem to have a strong ``inclination'' precisely for
that interval! Such is the case for fully developed turbulence in Couette-Taylor experiments \cite{turbulentbeck}
(with $1 \leq q \leq 1.2$ for Newtonian turbulence and $q \simeq 1.5$ for Lagrangian turbulence),
 {\it Hydra viridissima} \cite{hydra} ($q \simeq 1.5$), electron-positron annihilation \cite{bediaga} 
($1 \leq q \lesssim 1.3$), cosmic rays \cite{cosmic} ($q \simeq 1.2$), and finance \cite{economics3} 
($q \simeq 1.4-1.5$),
among others. There are, however, other situations in nature for which the case is not so clear-cut,
although still distinguishable by, e.g., a careful analysis of the behavior near the origin.
See Fig. 6 for comparison.

In the specific case of high-frequency financial observations, other studies
\cite{ple99,gop00} support the facts that (a) variances are finite and
(b) the exponent in the power-law tails lies outside the stable L{\'e}vy interval $2 > \gamma > 0$.  
These observations, together with the obvious presence of long tails, clearly demonstrate
the inadequacy of CLT distributions (L{\'e}vy or Gaussian) to describe high-frequency 
returns. In contrast, the use of {\it q}-Gaussians with $1 < q < 5/3$ is consistent both
with finite variances and with the presence of temporal autocorrelations (or
memory effects) in the dynamics of financial systems.

 \section*{Acknowledgments}
 One of us (CT) is very pleased to dedicate his invited talk to his old friend Idahlia Stanley. Two of us (CT and LB) thank Prof. H.E. Stanley and the organizers for their invitation to participate at the Bali meeting. This work has been partially supported by PRONEX/MCT, CNPq, and FAPERJ (Brazilian agencies).


 \vspace*{1cm}
{\bf Captions for figures} \\[2mm] \noindent

{\bf Figure 1:} {\small   
Implied volatilities as a function of the strike price for call options on
JY currency futures, traded on May 16 2002, with 147 days left to expiration. 
In this example the current price of a contract on Japanese futures is \$ 79,
 and the risk-free rate of return is 5.5 \%.
Circles correspond to  volatilities implied by the market, whereas triangles
correspond to volatilities implied by our model with $q=1.4$ and $\sigma= 10.2\%$. The dotted line is a guide to the eye.
 }
 \\[2mm] \noindent

 {\bf Figure 2:} {\small
 Empirical distributions (points) and {\it q}-Gaussians (solid lines) for normalized returns
 of the 10 top-volume stocks in the NYSE in 2001. The dotted line is the Gaussian
 distribution. The 2- and 3-min curves are moved vertically for display purposes.
 }
\\[2mm] \noindent

{\bf Figure 3:} {\small 
 Same as Fig. 2, but for the NASDAQ.}
 \\[2mm] \noindent

 {\bf Figure 4:} {\small
 Empirical distributions (points) and {\it q}-exponential-like fits (solid lines) for
 normalized volumes of the 10 top-volume stocks in the NYSE in 2001.
 }
\\[2mm] \noindent

 {\bf Figure 5:} {\small
 Same as Fig. 4, but for the NASDAQ.
 }
\\[2mm] \noindent

 {\bf Figure 6:} {\small
 L\'evy distributions $L_\gamma(x)\equiv \frac{1}{2\pi} {\displaystyle \int^\infty_{-\infty}} {\rm d}k \cos{k x} \, 
 {\rm e}^{-\alpha|k|^\gamma}$, with $0 < \gamma < 2$ and $\alpha>0$   (black curves), 
 and $q$-Gaussians $P_q(x)\equiv [1-(1-q)\beta x^2]^{1/(1-q)}/Z_q$, with  $5/3 <q <3$, $\beta>0$ and 
 $Z_q=\sqrt{\frac{\pi}{\beta(q-1)}}  \Gamma(\frac{3-q}{2(q-1)}) / \Gamma(\frac{1}{q-1})$ (red curves). 
 Parameters $(q,\gamma)$ are related through $q=\frac{\gamma+3}{\gamma+1}$
 so that the tails of both distributions decay with the same power-law exponent. 
 Without loss of generality, we have taken $\beta=1$ which corresponds to a simple rescaling; 
 $\alpha$ was chosen such that $P_q(0)=L_\gamma(0)$.
 }


\begin{thebibliography}{10}
 %
 \bibitem{egdcohen} E.G.D. Cohen, Physica A {\bf 305}, 19 (2002).


 \bibitem{takens}F. Takens, in {\it Structures in dynamics - Finite 
 dimensional deterministic studies}, eds. H.W. Broer, F. Dumortier, S.J. 
 van Strien and F. Takens (North-Holland, Amsterdam, 1991), page 253.


 \bibitem{krylov}N. Krylov, Nature {\bf 153}, 709 (1944). For full details see N.S. Krylov, {\it Works on the Foundations of Statistical Physics}, translated by A.B. Migdal, Ya. G. Sinai and Yu. L. Zeeman, Princeton Series in Physics (Princeton University Press, Princeton, 1979).


 \bibitem{gell-mann}M. Gell-Mann, {\it The Quark and the Jaguar: Adventures in the Simple and the Complex} (W.H. Freeman, New York, 1999). 


 \bibitem{baranger} M. Baranger, Physica A {\bf 305}, 27 (2002).


 \bibitem{turbulentbeck}C. Beck, G. S. Lewis and H. L. Swinney,
 Phys. Rev. E {\bf 63}, 035303 (2001); C. Beck, Phys. Rev. Lett. {\bf 87}, 180601 (2001).


 \bibitem{turbulentarimitsu}T. Arimitsu and N. Arimitsu, Physica A {\bf 305}, 218 (2002).


 \bibitem{bediaga}I. Bediaga, E. M. F. Curado and J. Miranda, Physica A {\bf 286}, 156 (2000).


 \bibitem{cosmic}C. Tsallis, J.C. Anjos and E.P. Borges, {\it Fluxes of cosmic rays: A delicately balanced anomalous stationary state}, astro-ph/0203258 (2002).


 \bibitem{economics1}C. Anteneodo, C. Tsallis and A.S. Martinez, Europhys. Lett. {\bf 59}, 635 (2002).


 \bibitem{economics2}L. Borland, Phys. Rev. Lett. {\bf 89}, 098701 (2002);  Quantitative Finance {\bf 2}, 415 (2002). 


 \bibitem{economics3}R. Osorio, L. Borland and C. Tsallis, in {\it Nonextensive Entropy - Interdisciplinary Applications}, M. Gell-Mann and C. Tsallis, eds. (Oxford University Press, 2003), in preparation; see also F. Michael and M.D. Johnson, 
 {\it Financial market dynamics}, Physica A (2002), in press. 


 \bibitem{hydra}A. Upadhyaya, J.-P. Rieu, J.A. Glazier and Y. Sawada, Physica A {\bf 293}, 549 (2001).


 \bibitem{ademir}J. A. S. de Lima, R. Silva and A. R. Plastino, Phys. Rev. Lett. {\bf 86}, 2938 (2001).


 \bibitem{chaos1}C. Tsallis, A.R. Plastino and W.-M. Zheng, Chaos, Solitons \& Fractals {\bf 8}, 885 (1997); U.M.S. Costa, M.L. Lyra, A.R. Plastino and C. Tsallis,
 Phys. Rev. E {\bf 56}, 245 (1997); M.L. Lyra and C. Tsallis, Phys. Rev. Lett. {\bf 80}, 53 (1998); U. Tirnakli, C. Tsallis and M.L. Lyra, Eur. Phys. J. B {\bf 11}, 309 (1999); V. Latora, M. Baranger, A. Rapisarda, C. Tsallis, Phys. Lett. A {\bf 273}, 97 (2000); F.A.B.F. de Moura, U. Tirnakli, M.L. Lyra, Phys. Rev. E {\bf 62}, 6361 (2000); U. Tirnakli, G. F. J. Ananos, C. Tsallis, Phys. Lett. A {\bf 289}, 51 (2001); H. P. de Oliveira, I. D. Soares and E. V. Tonini, Physica A {\bf 295}, 348 (2001); F. Baldovin and A. Robledo, Europhys. Lett. {\bf 60}, 518 (2002);  F. Baldovin and A. Robledo, Phys. Rev. E {\bf 66}, 045104(R) (2002); E.P. Borges, C. Tsallis, G.F.J. Ananos and P.M.C. Oliveira, Phys. Rev. Lett. {\bf 89}, 254103 (2002); U. Tirnakli, Physica A {\bf 305}, 119 (2002); U. Tirnakli, Phys. Rev. E {\bf 66}, 066212 (2002). 


 \bibitem{chaos2}Y. Weinstein, S. Lloyd and C. Tsallis, Phys. Rev. Lett. {\bf 89}, 214101 (2002).


 \bibitem{entanglement}S. Abe and A.K. Rajagopal, Physica A {\bf 289}, 157 (2001), C. Tsallis; S. Lloyd and M. Baranger, Phys. Rev. A {\bf 63}, 042104 (2001); C. Tsallis, P.W. Lamberti and D. Prato, Physica A {\bf 295}, 158 (2001); F.C. Alcaraz and C. Tsallis, Phys. Lett. A {\bf 301}, 105 (2002); C. Tsallis, D. Prato and C. Anteneodo, Eur. Phys. J. B {\bf 29}, 605  (2002); J. Batle, A.R. Plastino, M. Casas and A. Plastino, {\it Conditional $q$-entropies and quantum separability: A numerical exploration}, quant-ph/0207129 (2002).


 \bibitem{longrange}C. Anteneodo and C. Tsallis, Phys. Rev. Lett {\bf 80}, 5313 (1998); V. Latora, A. Rapisarda and C. Tsallis, Phys. Rev. E {\bf 64}, 056134 (2001); 
 A. Campa, A. Giansanti and D. Moroni, in {\it Non Extensive Statistical Mechanics and Physical Applications}, eds. G. Kaniadakis, M. Lissia and A. Rapisarda,  Physica A {\bf 305}, 137 (2002); B.J.C. Cabral and C. Tsallis, Phys. Rev. E {\bf 66}, 065101(R) (2002); E.P. Borges and C. Tsallis, in {\it Non Extensive Statistical Mechanics and Physical Applications}, eds. G. Kaniadakis, M. Lissia and A. Rapisarda,  Physica A {\bf 305}, 148 (2002); A. Campa, A. Giansanti, D. Moroni and C. Tsallis, Phys. Lett. A {\bf 286}, 251 (2001); M.-C. Firpo and S. Ruffo, J. Phys. A {\bf 34}, L511 (2001); C. Anteneodo and R.O. Vallejos, Phys. Rev. E  {\bf 65}, 016210 (2002); R.O. Vallejos and C. Anteneodo, Phys. Rev. E {\bf 66}, 021110 (2002); 
 M.A. Montemurro, F. Tamarit and C. Anteneodo, {\it Aging in an infinite-range Hamiltonian system of coupled rotators}, Phys. Rev. E (2003), in press.

 \bibitem{internet}S. Abe and N. Suzuki, {\it Itineration of the Internet over nonequilibrium stationary states in Tsallis statistics}, Phys. Rev. E (2002), in press.


 \bibitem{tsallis1}C. Tsallis, J. Stat. Phys. {\bf 52}, 479 (1988).


 \bibitem{tsallis2}E.M.F. Curado and C. Tsallis, J. Phys. A: Math. Gen. {\bf 24}, L69 (1991) [Corrigenda: 
 {\bf 24}, 3187 (1991) and {\bf 25}, 1019 (1992)]; C. Tsallis, R.S. Mendes and A.R. Plastino, Physica A {\bf 261}, 534 (1998).


 \bibitem{tsallis3}S.R.A. Salinas and C. Tsallis, eds., {\it Nonextensive Statistical Mechanics and Thermodynamics}, Braz. J. Phys. {\bf 29}, No. 1 (1999);
 S. Abe and Y. Okamoto, eds., {\it Nonextensive Statistical Mechanics and its Applications}, Series {\it Lecture Notes in Physics} (Springer-Verlag, Berlin, 2001); G. Kaniadakis, M. Lissia and A. Rapisarda, eds., {\it Non Extensive Statistical Mechanics and Physical Applications}, Physica A {\bf 305}, No 1/2 (Elsevier, Amsterdam, 2002); M. Gell-Mann and C. Tsallis, eds., {\it Nonextensive Entropy - Interdisciplinary Applications} (Oxford University Press, 2003), in preparation; H.L. Swinney and C. Tsallis, eds.,  {\it Anomalous Distributions, Nonlinear Dynamics, and Nonextensivity}, Physica D (2003), in preparation. An updated bibliography can be found at the web site
 http://tsallis.cat.cbpf.br/biblio.htm


 \bibitem{stanley}R. N. Mantegna and H. E. Stanley, Nature {\bf 376}, 46 (1995);  R. N. Mantegna and H. E. Stanley, {\it Introduction to Econophysics: Correlations \& Complexity in Finance} (Cambridge University Press, Cambridge, 2000); H. E. Stanley, L. A. N. Amaral, S. V. Buldyrev, P. Gopikrishnan, V. Plerou, and M. A. Salinger, in Proc. Natl. Acad. Sci. Arthur M. Sackler Colloquium, {\it Self-Organized Complexity in the Physical, Biological, and Social
 Sciences} (23--24 March 2001, Beckman Center, Irvine, CA), Proc. Natl. Acad. Sci. {\bf 99}, Supp.1: 2561 (2002).


 \bibitem{aversion}D. Kahneman and A. Tversky, Econometrica {\bf 47}, 263 (1979); A. Tversky and D. Kahneman, Journal of Risk and Uncertainty {\bf 5}, 297 (1992); A. Tversky and P. Wakker, Econometrica {\bf 63}, 1255 (1995); A. Tversky and C.R. Fox, Psychological Review {\bf 102} (1995) 269; R. Gonzalez and G. Wu, Cognitive Psychology {\bf 38}, 129 (1999).



 \bibitem{pdependent}  L. Borland, Phys. Rev. E {\bf 57}, 6634 (1998).
 
 
 
\bibitem{liu99}
Liu, Y., P. Gopikrishnan, P. Cizeau, M. Meyer, C.~K. Peng, and H.~
E. Stanley,  Phys. Rev. E {\bf 60}, 1390 (1999).

\bibitem{ple99}
Plerou, V., P. Gopikrishnan, L.~A.~N. Amaral, M. Meyer, and H.~E.
Stanley, Phys. Rev. E {\bf 60}, 6519 (1999).

\bibitem{gop00}
Gopikrishnan, P., V. Plerou, Y. Liu, L.~A.~N. Amaral, X. Gabaix,
and H.~E. Stanley, Physica A {\bf 287}, 362 (2000).

\bibitem{gop00volu}
Gopikrishnan, P., V. Plerou, X. Gabaix, and H.~E. Stanley,
Phys. Rev. E {\bf 62}, R4493 (2000).

\bibitem{abde01}
Andersen, T.~G., T. Bollersev, F.~X. Diebold, and H. Ebens,  J. Financial
Economics {\bf 63}, 43 (2001).



 \bibitem{nonlinearFP}A.R. Plastino and A. Plastino, Physica A  {\bf 222}, 347 (1995); 
 C. Tsallis and D.J. Bukman, Phys. Rev. E {\bf 54}, R2197 (1996); 
 C. Giordano, A.R. Plastino, M. Casas and A. Plastino, Eur. Phys. J. B {\bf 22}, 361 (2001); 
 A. Compte and D. Jou, J. Phys. A {\bf  29}, 4321 (1996); A.R. Plastino, M. Casas and A. Plastino, Physica A {\bf 280}, 289 (2000); 
 C. Tsallis and E.K. Lenzi, in {\it Strange Kinetics}, eds. R. Hilfer et al, Chem. Phys.  {\bf 284}, 341 (2002) [Erratum (2002)]; 
 E.K. Lenzi, L.C. Malacarne, R.S. Mendes and I.T. Pedron, {\it Anomalous diffusion, nonlinear fractional Fokker-Planck equation 
 and solutions}, Physica A (2003), in press [cond-mat/0208332]; E.K. Lenzi, C. Anteneodo and L. Borland, Phys. Rev. E {\bf 63}, 051109 (2001); 
 E.M.F. Curado and F.D. Nobre, {\it Derivation of nolinear Fokker-Planck equations by means of approximations to the master 
 equation}, Phys. Rev. E (2003), in press; C. Anteneodo and C. Tsallis, {\it Multiplicative noise: A mechanism leading to 
 nonextensive statistical mechanics}, cond-mat/0205314 (2002).


 \bibitem{bologna}M. Bologna, C. Tsallis and P. Grigolini, Phys. Rev. E {\bf 62}, 2213 (2000).

 \bibitem{zanettemontemurro}D.H. Zanette and M.A. Montemurro, {\it Thermal measurements of stationary nonequilibrium systems: 
 A test for generalized thermostatistics}, cond-mat/0212327 (2002).


 \end{thebibliography}
 
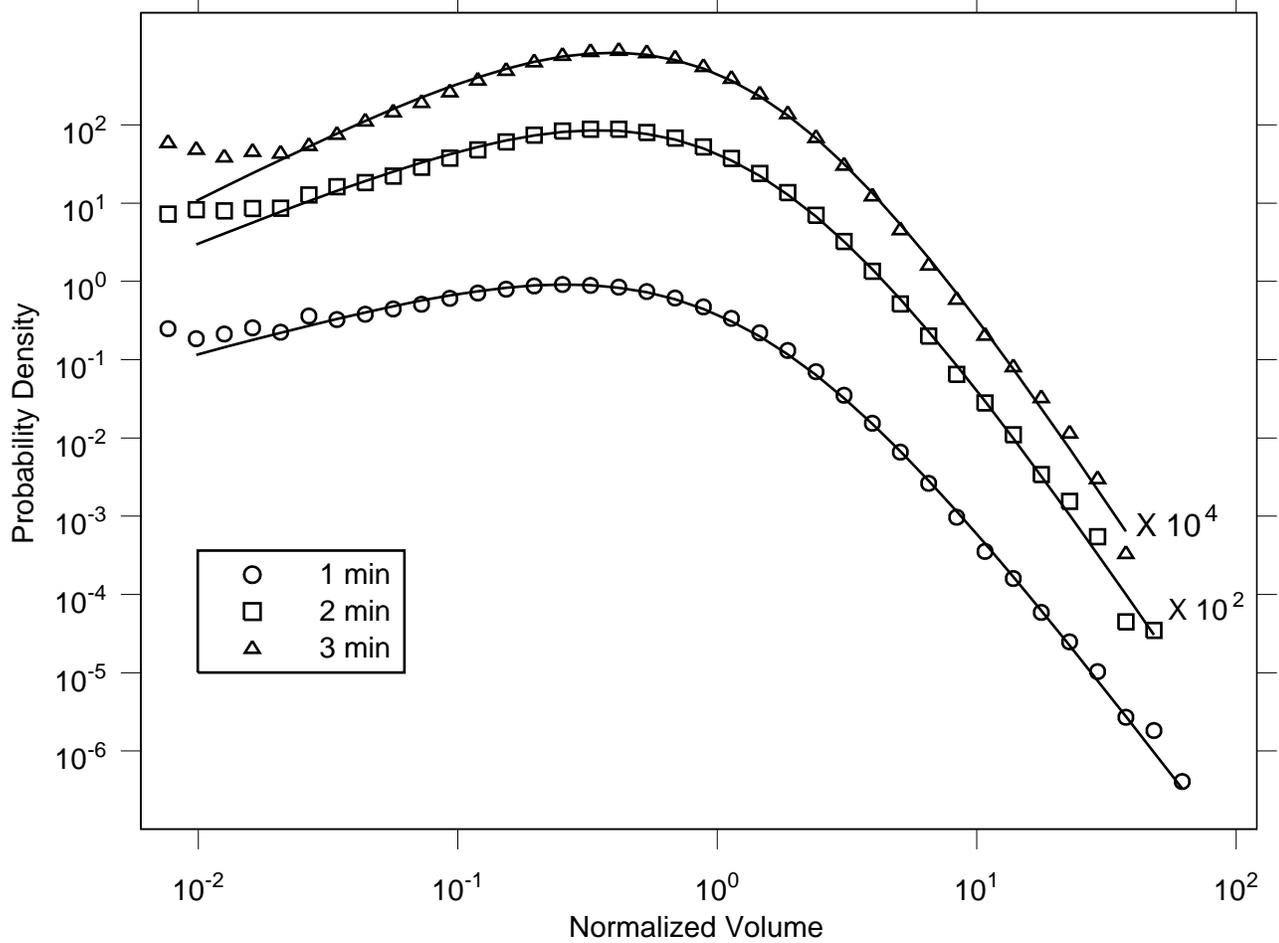
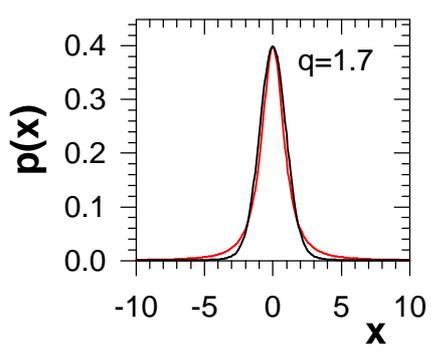
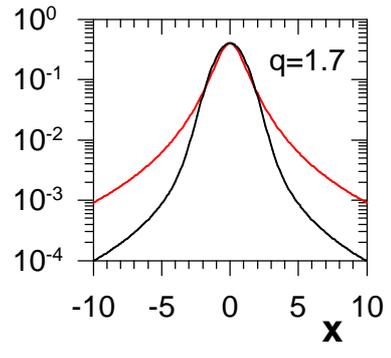
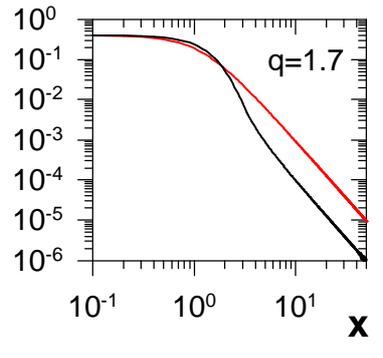
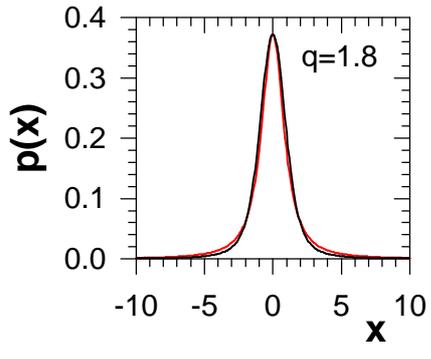
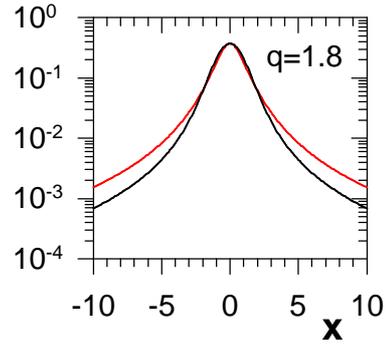
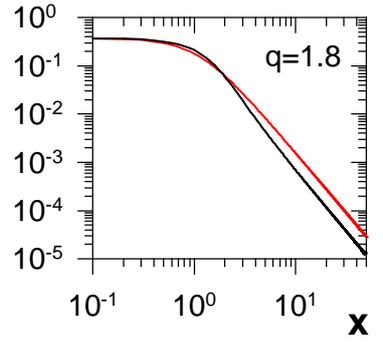
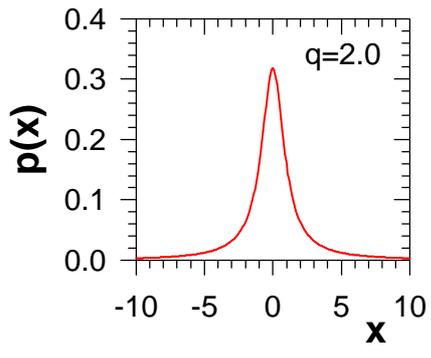
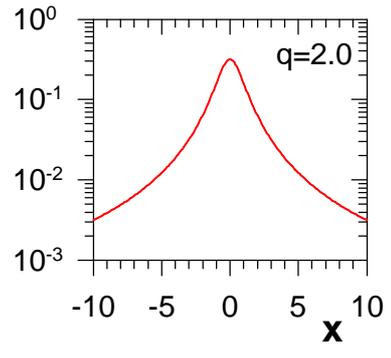
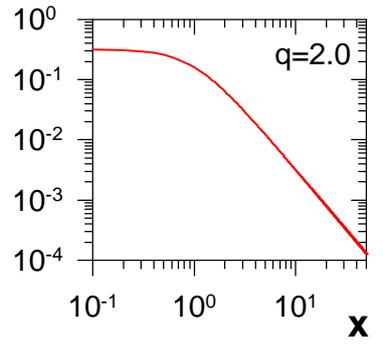
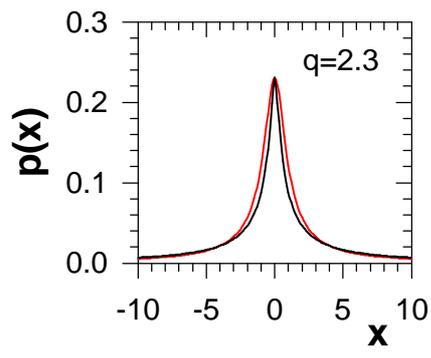
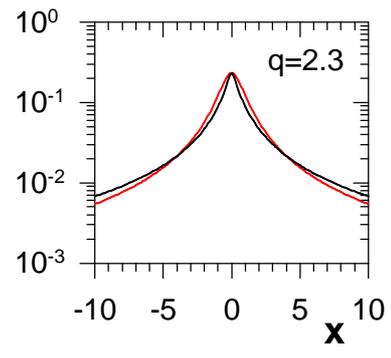
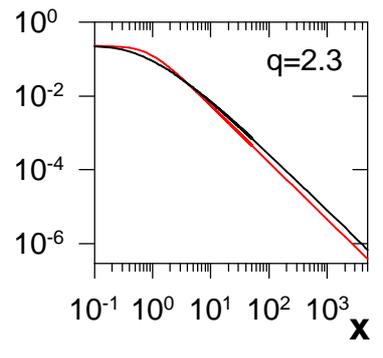
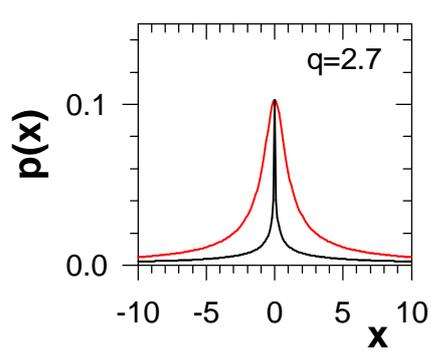
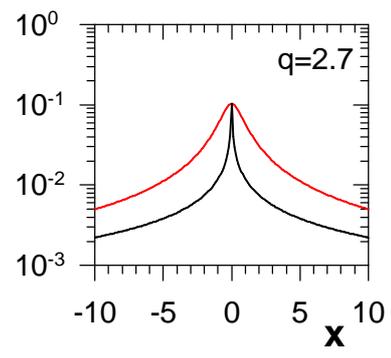
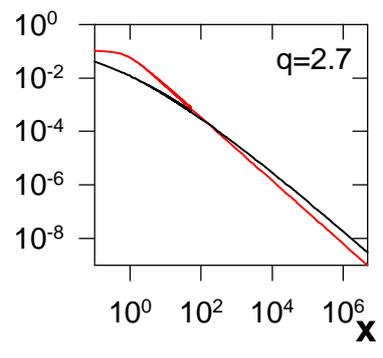
\end{document}